\documentclass[12pt]{article}
\usepackage{a4}
\usepackage{citesort}
\usepackage{amsmath}
\usepackage{amssymb}
\usepackage{latexsym}
\def \d{{\mathrm{d}}}

\def \pd{\partial}

\def \iner{\rfloor}
\def \hodge{{}^\star}

\def \tl#1{\overset{\kern 1pt\circ}{#1}}
\def \TL#1{\overset{\kern -3pt \circ}{#1}}
\def \TLL#1{\overset{\kern -7pt \circ}{#1}}

\def\vt{\vartheta}
\def \LL{{\cal{L}}}

\begin{document}

\title{{\bf On the fundamentals of the three-dimensional
translation gauge theory of dislocations}}
\author{Markus Lazar~$^\text{a,b,}$\footnote{E-mail:
    lazar@fkp.tu-darmstadt.de (M.~Lazar).}$$
\\ \\
${}^\text{a}$
        Emmy Noether Research Group,\\
        Department of Physics,\\
        Darmstadt University of Technology,\\
        Hochschulstr. 6,\\
        64289 Darmstadt, Germany\\
${}^\text{b}$ 
Department of Physics,\\
Michigan Technological University,\\
Houghton, MI 49931, USA
}

\date{\today}    
        
\maketitle
\begin{abstract}
We propose a dynamic version of the three-dimensional translation 
gauge theory of dislocations. 
In our approach, we use the notions of the dislocation density and dislocation 
current tensors 
as translational field strengths and the corresponding response quantities 
(pseudomoment stress, dislocation momentum flux). 
We derive a closed system of field equations in a
very elegant quasi-Maxwellian form 
as equations of motion for dislocations.
In this framework, 
the dynamical Peach-Koehler force density is derived as well.
Finally, the similarities and the differences between the Maxwell field theory
and the dislocation gauge theory are presented.
\\

\end{abstract}
\vspace*{5mm}
\noindent
\section{Introduction}
In the last years, there has been a growing interest in continuum theories of dislocations.
This development has been driven by the explanation of size effects in small-scale 
structures and by physically based plasticity theories of dislocations 
(see, e.g.,~\cite{Plast}). 
On the other hand,
it has been known for a long time that fast moving dislocations 
exhibit typical properties of moving particles and electromagnetic 
fields~\cite{frank,eshelby,leibfried}. 
The stress field of a moving dislocation is longitudinally contracted.
For that reason, the analogy between the theory of dislocations
and the Maxwell theory of electromagnetic fields is often discussed
in the literature~\cite{kroener55,kroener81,hollaender,bovet,mistura95}. 
Kr{\"o}ner~\cite{kroener55} 
proposed an analogy between the dislocation theory and the theory of the 
magnetic field of distributions of stationary electric currents.
Other authors~\cite{bovet,mistura95} suggested an analogy between the 
deformation and the magnetic fields and between the velocity and the electric
fields. 
Also some authors~\cite{goleb,edelen80} used the analogy between the magnetic field 
and the dislocation density and between the electric field and the dislocation 
current density. Nevertheless, they did not use the concepts of excitations 
in dislocation theory which are necessary to obtain a complete theory 
with a closed system of equations of motion
analogous to the electromagnetic field theory. 
Already, Schaefer~\cite{Schaefer} 
pointed out that constitutive equations are missing in the classical dislocation theory.

A dynamical theory of dislocations was formally derived in \cite{kadic,Edelen88}. 
We would like to mention that the Lagrangian of dislocations 
that they proposed contains only 
one material constant 
for the dislocation density tensor as well as for the dislocation current tensor.
An improved and more realistic 
dislocation model has been formulated in~\cite{Lazar00,Lazar02,LA08,LA09}.
In~\cite{LA09}, the static solutions of screw and edge dislocations were given.
The linear version of a dynamic dislocation gauge theory was formulated in~\cite{LA08}
and successfully applied to a moving screw dislocation in the subsonic as well as in the
supersonic regimes~\cite{Lazar09}. The nonuniformly motion of a dislocation 
has been investigated in~\cite{Lazar10}.
Recently, Lazar and Hehl~\cite{LH09} have investigated Cartan's spiral staircase in 
the gauge theory of dislocations and they have shown that such a configuration
arises naturally as a solution of the three-dimensional theory of dislocations.

The aim of this paper is to develop a dynamic theory of dislocations
more systematically 
which makes use of the concepts of field strengths, excitations and constitutive 
relations. 
This theory is a kind of axiomatic field theory of dislocations. 
The nature of such a general dislocation theory is geometrically nonlinear.
We are using the analogies to the 
axiomatic Maxwell theory given by Hehl and Obukhov~\cite{HO01}. 
The language of differential forms~\cite{frankel}
is used as a powerful mathematical tool for overcoming structural decencies.
Moreover, 
the framework of metric affine gauge theory~(MAG) given
by Hehl et al.~\cite{Hehl95} is also used 
in order to obtain a clear description of dislocation systems.
The purpose of the paper is to give the differential geometric structure of the theory
and not to discuss solutions of the field equations. 
For solutions of dislocations in the framework of dislocation gauge theory we refer 
to~\cite{Lazar02,LA09,Lazar09,Lazar10,LH09}.


\section{Elastodynamics}
In finite elasticity theory~(see, e.g.,~\cite{ogden,marsden,Maugin}), 
the material body is identified with a 
three-dimensional manifold ${\cal M}^3$ which is embedded into the 
three-dimensional Euclidean space ${\mathbb R}^3$. We distinguish between  the 
material or the  final coordinates of ${\cal M}^3$, $a,b,c,\ldots=1,2,3$,  
and the (holonomic) Cartesian coordinates of the reference system  
(defect-free or ideal reference system) ${\mathbb R}^3$, $i,j,k,\ldots=1,2,3$.
A {\it deformation} of ${\mathbb R}^3$ is a mapping ${\boldsymbol\xi}:\ 
{\mathbb R}^3\longrightarrow{\cal M}^3$.
A time-dependent family of deformation is called motion of ${\cal M}^3$.
The {\it distortion} 1-form is defined by
\begin{align}
\label{dist1}
\vartheta^a=B^a{}_i\,\d x^i=\d\xi^a
\end{align}
and is identified with a coframe. In elasticity, $B^a{}_i=\pd \xi^a/\pd x^i$ 
is called the {\it deformation gradient}, which is a so-called 
two-point tensor field because it is defined on two configurations or
bases. 
$\vt^a$ and $\xi^a$ have the dimensions: $[\vt^a]={\text{length}}$ and 
$[\xi^a ]={\text{length}}$.
Here $\d$ is the three-dimensional exterior derivative. 
If the coframe (or distortion) 1-form has the property that
\begin{align}
\label{cc1}
\d\vartheta^a=0\, ,
\end{align}
it said to be holonomic or compatible. 
Therefore, Eq.~(\ref{cc1}) is a compatibility condition for $\vt^a$.
The frame field 
\begin{align}
e_a= e_a{}^i \, \pd_i
\end{align}
is dual to the coframe~(\ref{dist1}) such that:
\begin{align}
e_a\iner\vartheta^b=e_a^{\ i}\,B^b_{\ i}=\delta_a^b\, .
\end{align}
Here $\iner$ denotes the interior product.
The frame $e_a$ has the dimension: $[e_a]=1/\text{length}$.

A $\vartheta$-basis for 0-, 1-, 2-, and 3-forms is
$\left\{1,\vt^a,\,\vt^{ab}:=\vt^a\wedge\vt^b,\,\vt^{abc}:=
\vt^a\wedge\vt^b\wedge\vt^c\right\}$, the Hodge dual $\eta$-basis for
3-,2-,1-, and 0-forms is specified by
\begin{align}
  \eta&:=\hodge 1=\frac{1}{3!}\, \eta_{abc}\, \vt^{abc}\,,\\
  \eta_a&:=\hodge\vt_a=e_a\iner\eta=\frac{1}{2}\,
  \eta_{abc}\, \vt^{bc}\,,\\
  \eta_{ab}&:=\hodge(\vt_{ab})=e_b\iner\eta_a= \eta_{abc}\, \vt^c\,,\\
  \eta_{abc}&:=\hodge(\vt_{abc})=e_c\rfloor\eta_{ab}  \,,
\end{align}
where $\hodge$ denotes the Hodge star,
$\wedge$ is the exterior product and
$\eta_{abc}={{\text{det}}\, (B^d_{\ i})}\,
\epsilon_{abc}$, with $\epsilon_{abc}$ as the totally
antisymmetric Levi-Civita symbol with $\pm1,0$.


Simultaneously, 
the {\it physical velocity} 0-form of the motion of the material continuum 
is given by
\begin{align}
\label{v-el}
v^a=\frac{\pd \xi^a}{\pd t}\equiv\dot\xi^a.
\end{align}
It is a vector-valued 0-form.
$v^a$ describes the velocity of material points of the continuum
and it has the dimension: $[v^a]=\text{length}/\text{time}$.
The time-dependent distortion $\vt^a$ and the velocity field $v^a$ have to satisfy
the following kinematical compatibility condition:
\begin{align}
\label{cc2}
\d v^a -\dot\vt^a =0\, .
\end{align}
In elasticity, 
this is the kinematical compatibility condition between the deformation gradient 
and the associated velocity field.

The {\it right Cauchy-Green tensor} $G$
is defined as the metric of the final state 
\begin{align}
G=g_{ab}\,\vartheta^a\otimes\vartheta^b
=g_{ab}\, B^a_{\ i}B^b_{\ j}\,\d x^i\otimes\d x^j
=g_{ij}\,\d x^i\otimes\d x^j.
\end{align}
If the coframe is orthonormal,
then it reads
\begin{align}
G=\delta_{ab}\,\vartheta^a\otimes\vartheta^b
=\delta_{ab}\, B^a_{\ i}B^b_{\ j}\,\d x^i\otimes\d x^j
\end{align}
with $\delta_{ab}=\text{diag}(+++)$.
The {\it Lagrangian strain} tensor is given by
\begin{align}
2E=G-1=(g_{ij}-\delta_{ij})\,\d x^i\otimes\d x^j.
\end{align}
It measures the change of the metric between the undeformed and the deformed 
state.

The question of the response quantities (elastic excitations) is now of interest.
This question is closely connected with the elastic field Lagrangian. 
In the continuum approach the elastic Lagrangian depends continuously on the elastic 
distortion and velocity. 
Thus, the {\it elastic Lagrangian} density 3-form is given by
\begin{align}
{\cal L}_{\mathrm{el}}={\cal L}_{\mathrm{el}}(v^a,\vt^a)\, .
\end{align}
As usual,
the elastic Lagrangian density can be given in terms of kinetic and potential energy
densities
\begin{align}
\label{el_lag1}
{\cal L}_{\mathrm{el}}=T_{\rm k}-W\, ,
\end{align}
where $T_{\rm k}$ is the {\it kinetic energy density} three-form
and $W$ denotes the {\it elastic potential energy density} 3-form (or distortion energy).
The elastic potential energy density is a measure of the energy stored in the material as a 
result of elastic deformation.
The excitation with respect to the physical velocity 0-form is the elastic 
{\it momentum} 3-form 
\begin{align}
p^{\rm }_a:=\frac{\pd {\cal L}_{\rm el}}{\pd v^a}\, .
\end{align}
It is a covector valued 3-form
\begin{align}
p_a =\frac{1}{3!}\, p_{aijk}\, \d x^i\wedge \d x^j\wedge\d x^k\, ,
\end{align}
which has the dimension: $[p_a]={\text{momentum}}
\stackrel{\rm SI}{=}{\text{N\,s}}$,
and $[p_{aijk}]=\text{momentum}/(\text{length})^3$.
The usual momentum vector (covector valued 0-form), 
known form elasticity theory, is given by:
$P_a=(1/3!)\eta^{ijk} p_{aijk}$. 
In elasticity, the linear and isotropic constitutive law for the momentum vector 
is of the form: $P_a=\rho g_{ab} v^b$, where $\rho$ denotes the mass density.

The elastic {\it force stress} 2-form is the excitation quantity 
relative to the
distortion 1-form and is defined by 
\begin{align}
\label{stress}
\Sigma^{\rm }_a:=\frac{\delta {\cal L}_{\rm el}}{\delta \vartheta^a}
=\frac{\pd {\cal L}_{\rm el}}{\pd \vartheta^a}\, .
\end{align}
Eq.~(\ref{stress}) is the general constitutive relation for nonlinear 
elasticity.
$\Sigma_a$ is a covector valued 2-form 
\begin{align}
\Sigma^{\rm }_a=\frac{1}{2}\, \Sigma_{aij}\, \d x^i\wedge\d x^j\, .
\end{align}
It has the dimension: $[\Sigma_a]={\text{force}}
\stackrel{\rm SI}{=}{\text{N}}$, and 
for the components: 
$[\Sigma_{aij}]={\text{force}/(\text{length})^2}=\text{stress}
\stackrel{\rm SI}{=}\text{P}$.
We recognize that the force stress has 9 independent components.
Therefore, the first Piola-Kirchhoff stress tensor (a two-point tensor) 
is represented by: $\Sigma_{a}{}^k=(1/2) \eta^{ijk} \Sigma_{aij}$.
In elasticity theory, the first Piola-Kirchhoff stress tensor reads:
$\Sigma_{a}{}^k=e_{aj}\sigma^{jk}$, where the Cauchy stress tensor
is given by the generalized Hooke law 
$\sigma^{jk}=C^{jkmn} E_{mn}$ with the elasticity tensor $C^{jkmn}$.


In elastodynamics, 
the Euler-Lagrange equations give the following equation of motion 
\begin{align}
\frac{\delta\LL_{\rm {el}}}{\delta \xi^a}&\equiv
\frac{\pd \LL_{\rm {el}}}{\pd \xi^a}-\d\, \frac{\pd\LL_{\rm {el}}}{\pd \d \xi^a}
-\pd_t\, \frac{\pd\LL_{\rm {el}}}{\pd \dot \xi^a}=0
\end{align}
and in terms of the stress 2-form and momentum 3-form they are of the form
\begin{align}
\dot{p}_a+\d \Sigma_a=0\, .
\end{align}

The {\it elastic energy} density ${\cal E}_{\rm el}$ is defined in the framework of field theory 
as the Hamiltonian of the elastic system
\begin{align}
{\cal E}_{\rm el}:=p^{\rm }_a v^a-{\cal L}_{\rm el}
                  =T_{\rm k}+W  .
\end{align}

\section{T(3)-gauge theory of dislocations}

In this section, we discuss the three-dimensional translation gauge theory of
dislocations. 
In this approach, we consider the translation group, $T(3)$, as a gauge group.
We assume that all fields depend on the space and time variables.
The $T(3)$-transformation acts on $\xi^a$  as 
a gauge transformation in the following way
\begin{align}
\label{xi-gt}
\xi^a\longrightarrow \xi^a-\tau^a(x,t)\, ,
\end{align}
where $\tau^a(x,t)$ are local and time-dependent translations.
If we do so, 
the invariance of the compatible distortion (\ref{dist1}) and the 
material velocity~(\ref{v-el})
are lost under the local $T(3)$-transformations~(\ref{xi-gt}).
In order to compensate the invariance violating terms, we have to introduce 
gauge potentials, a vector-valued 1-form $\phi^a=\phi^a{}_i\, \d x^i$
and a vector valued 0-form $\varphi^a$
transforming under the local transformations in a suitable way:
\begin{align}
&\phi^a\longrightarrow\phi^a+\d\tau^a(x,t)\\
&\varphi^a\longrightarrow\varphi^a+\dot{\tau}^a(x,t)\, .
\end{align}
Thus, $\phi^a$ and $\varphi^a$ are the translational gauge potentials
of the dynamical $T(3)$-gauge theory.
Since the elastic distortion and the material velocity are 
state quantities in the field theory of dislocations, they have to be 
gauge-invariant.
Now we redefine the elastic distortion and the material velocity 
in a gauge-invariant form as follows:
\begin{align}
\label{dist3}
\vartheta^a&:=\d \xi^a+\phi^a\\
\label{vel3}
v^a&:=\dot \xi^a+\varphi^a\, .
\end{align}
Hence, $\xi^a$, $\phi^a$ and $\varphi^a$ appear in (\ref{dist3}) and 
(\ref{vel3}) always joined together in the translation-invariant combinations.
The gauge fields $\phi^a$ and $\varphi^a$ make the elastic distortion 
$\vartheta^a$ and the material velocity $v^a$ incompatible because they are not
any longer just a simple gradient and a time-derivative of $\xi^a$.

In \cite{Lazar00} we have seen that $\phi^a$ in Eq.~(\ref{dist3}) can be 
interpreted as the translational part of the generalized affine connection 
in a Weitzenb{\"o}ck space. 
The underlying geometrical structure of the theory is the 
affine tangent bundle $AM$.
At every point, the tangent space is replaced by the 
affine tangent space.
The translation group $T(3)$ acts on the affine space as an internal
symmetry. 
The field $\xi^a$ is also known
as Cartan's `radius vector' and determines  the `origin'
of the affine space~\cite{Hehl95}.
In the context of dynamical $T(3)$-gauge theory, two translational gauge 
potentials $\phi^a$ and $\varphi^a$ and one translational Goldstone
field $\xi^a$ occur and they are the canonical field quantities.
On the other hand, in the `classical' theory of defects 
(see, e.g.,~\cite{Kossecka}) fields like $\phi^a$ and $\varphi^a$ are called 
the negative plastic (or initial) 
distortion and velocity, respectively.

\subsection{Translational field strengths}

Because we have two translational gauge potentials $\phi^a$ and $\varphi^a$, 
we may define two translational field strengths in terms of these gauge potentials.
We introduce the well-known quantities of dislocation density and 
dislocation current as field strengths of the $T(3)$-gauge theory breaking the
compatibility conditions~(\ref{cc1}) and (\ref{cc2}).
The {\it dislocation density} 2-form $T^a$ is defined as the {\it object of
anholonomity} or {\it torsion} 2-form in a teleparallel space (Weitzenb{\"o}ck space).
The torsion 2-form (dislocation density 2-form) is defined in terms of the
translational gauge potential $\phi^a$
\begin{align}
T^a=\d\phi^a
\end{align}
and, alternatively, in terms of the anholonomic coframe
\begin{align}
T^a=\d \vartheta^a\, .
\end{align}
Thus, $T^a$ measures how much the gauge potential $\phi^a$ and the 
coframe $\vt^a$ fail to be holonomic or compatible.
The torsion is a vector valued 2-form 
\begin{align}
T^a=\frac{1}{2}\, T^a{}_{ij}\,\d x^i\wedge\d x^j\, , 
\end{align}
which has the dimension: $[T^a]={\text{length}}$. The torsion tensor
$T^a{}_{ij}$ has the dimension: $[T^a{}_{ij}]={\text{1/length}}$.
The field $T^a_{\ 12}$ measures the number of dislocation lines going in 
3-direction and having the Burgers vector $b^a$.  
$T^a$ can describe a continuous distribution of dislocations as well as 
single dislocations.
The usual dislocation density tensor~\cite{kroener81} 
is represented by: 
$\alpha_a{}^k=(1/2)\eta^{ijk} T^a{}_{ij}$.

Another translational field strength is the {\it dislocation current}.
The dislocation current 1-form is defined in terms of the 
translational gauge potentials $\varphi^a$ and $\phi^a$ as follows
\begin{align} 
\label{I1}
I^a=\d \varphi^a-\dot\phi^a
\end{align}
or in terms of the incompatible coframe and incompatible velocity as 
\begin{align} 
\label{I2}
I^a=\d v^a-\dot\vt^a\, .
\end{align}
In Eq.~(\ref{I2}), we see that the dislocation current is defined in terms of
the physical velocity gradient and the rate of the elastic distortion.
It is a vector valued 1-form
\begin{align}
I^a=I^a{}_{i}\,\d x^i\, ,
\end{align}
with the dimension: $[I^a]={\text{velocity}}
\stackrel{\rm SI}{=}\text{m}/\text{s}$. The 
dislocation current tensor has the dimension:
$[I^a{}_i]=1/{\text{time}}
\stackrel{\rm SI}{=}1/\text{s}$.
The translational field strengths $T^a$ and $I^a$  measure the 
violation of the two compatibility conditions ~(\ref{cc1}) and (\ref{cc2})
and, therefore, 
 how much the elastic distortion $\vt^a$ and the physical velocity $v^a$ 
are incompatible. 
The dislocation density $T^a$ and the dislocation current $I^a$ are state quantities
and they can be observed experimentally. The dislocation current is
the appropriate quantity for the description of dynamics of dislocations.
For the dynamical case, $I^a$ and $T^a$ carry the information
about the dislocation state of motion. 

The two translational field strengths have to satisfy the following Bianchi identities:
\begin{align}
\label{bianchi1}
\d T^a&=0\, ,\\
\label{bianchi2}
\d I^a+\dot T^a&=0\, .
\end{align}
Eq.~(\ref{bianchi1}) is the well-known conservation law of dislocations 
(continuity equation of the dislocation density) and 
(\ref{bianchi2}) is known as the continuity equation of the dislocation
current~\cite{kosevich,landau}.
Eq.~(\ref{bianchi1}) states that a dislocation line cannot end inside the body.
The evolution of $T^a$ is
determined by $I^a$ in a closed form.
If we integrate Eq.~(\ref{bianchi1}) over a three-dimensional
volume which contains dislocations and use the Stokes theorem,
we obtain the Burgers vector  by means of the Burgers circuit $\pd S$
\begin{align}
\label{burger}
\int_{S}T^a=\oint_{\pd S} \vt^a=b^a\, 
\end{align}
where $\pd S$ is the boundary of the surface $S$.
On the other hand, if we integrate Eq.~(\ref{bianchi2}) over a two-dimensional surface 
$S$ and we use the Stokes theorem, we get
the so-called `conservation law of the Burgers vector'
\begin{align}
\label{burger2}
\oint_{\pd S} I^a+\dot b^a=0\, .
\end{align}
The integral of Eq.~(\ref{burger2}) determines the flux of the Burgers
vector $b^a$ per unit time through the contour $\pd S$.
It states that the time change of the Burgers vector on
a surface $S$ is equal to the negative dislocation current over the contour $\pd S$ 
of the surface $S$. 

If the local translation is not time-dependent $\tau^a=\tau^a(x)$, then 
the physical velocity $v^a$ is compatible like in (\ref{v-el}) and one may 
use the gauge condition $\varphi^a=0$ in order to eliminate the gauge potential
$\varphi^a$. 
Such a gauge condition is called temporal gauge (or Weyl gauge) in gauge field 
theories. In this gauge, the dislocation current~(\ref{I1}) 
reads: $I^a=-\dot\phi^a$. 
Such a situation is used, e.g., in~\cite{kosevich,landau} if we identify the 
gauge potential $\phi^a$ with the negative plastic distortion. In this way, 
the dislocation current is related to the rate of plastic distortion.

\subsection{Gauge field momenta -- dislocation field excitations}

To complete the field theory of dislocations, we have to define the 
excitations with respect to the dislocation density and the 
dislocation current, respectively.
The dislocation Lagrangian density is of the form:
\begin{align}
\LL_{\rm {disl}}=\LL_{\rm {disl}}(v^a,\vt^a,I^a,T^a)\, .
\end{align}
Temporarily we will leave open the explicit form of $\LL_{\rm {disl}}$. Then
it is necessary to introduce the field momenta that are canonically conjugated to
the translational field strengths. These field momenta are called the dislocation
excitations.
The excitation with respect to the torsion 2-form is defined by
\begin{align}
\label{H-CR}
H_a:=-\frac{\pd{\cal L}_{\rm disl}}{\pd T^a}\, .
\end{align}
It is the specific response to $T^a$.
Another excitation is the 2-form $D_a$ 
\begin{align}
\label{D-CR}
D_a:=\frac{\pd{\cal L}_{\rm disl}}{\pd I^a}\, .
\end{align}
It is the response of the dislocation Lagrangian to $I^a$.
We may interpret $H_a$ and $D_a$ as
the {\it pseudomoment stress}
and the {\it dislocation momentum flux}, respectively, 
caused by the motion of dislocations.
Here, $H_a$ is a covector valued 1-form
\begin{align}
H_{a}=H_{ai}\, \d x^i \, .
\end{align}
The dimension of $H_a$ is: $[H_a]={\text{force}}$, and 
for its components: $[H_{ai}]={\text{force}/\text{length}}$.
$D_a$ is a covector valued 2-form
\begin{align}
D_a=\frac{1}{2}\, D_{aij}\, \d x^i\wedge\d x^j\, ,
\end{align}
which has the dimension:
$[D_a]={\text{momentum}}$, and 
$[D_{aij}]=\text{momentum}/(\text{length})^2$.
Eqs.~(\ref{H-CR}) and (\ref{D-CR}) are constitutive relations of the nonlinear
dislocation field theory.

In addition, 
we define the {\it dislocation stress} as a covector valued 2-form as
\begin{align}
\label{E-CR}
E_a:=\frac{\pd{\cal L}_{\rm disl}}{\pd \vt^a}
\end{align}
with the dimension $[E_a]={\text{force}}$,
and the {\it dislocation momentum} 3-form reads
\begin{align}
\label{p-CR}
\pi_a:=\frac{\pd{\cal L}_{\rm disl}}{\pd v^a}\, ,
\end{align}
which is a covector valued 3-form and it has the dimension:
$[\pi_a]={\text{momentum}}$.
Eqs.~(\ref{E-CR}) and (\ref{p-CR}) are valid
for the dislocation stress and the dislocation momentum in 
the nonlinear dislocation field theory.
The dislocation stress 2-form is explicitely given by
\begin{align}
\label{E}
E_a=e_a\iner \LL_{\text{disl}}+(e_a\iner T^b)\wedge H_b-(e_a\iner I^b) D_b
\end{align}
and the dislocation momentum 3-form reads
\begin{align}
\label{mom_disl2}
\pi_a=(e_a\iner T^b)\wedge D_b=-(e_a\iner D_b)\wedge T^b\, .
\end{align}
In continuum mechanics, (\ref{E}) and (\ref{mom_disl2}) are called the Eshelby
stress and the pseudomomentum (see, e.g.,~\cite{Eshelby,Maugin}).

\subsection{Dislocation field equations}

We are now ready to 
derive the Yang-Mills type field equations determining the dynamics of dislocations.
The total Lagrangian density is of the form
\begin{align}
\label{L}
{\cal L}={\cal L}_{\rm disl}+{\cal L}_{\rm el}\, .
\end{align}
The variation of the total Lagrangian with respect to the Goldstone field
$\xi^a$ and the translational gauge potentials
$\varphi^a$ and $\phi^a$
gives the Euler-Lagrange equations. According
to the extremal action principle, the field equations are found to be
\begin{align}
\label{FE-xi}
\frac{\delta\LL}{\delta \xi^a}&\equiv
\frac{\pd\LL}{\pd \xi^a}-\d\, \frac{\pd\LL}{\pd \d \xi^a}
-\pd_t\, \frac{\pd\LL}{\pd \dot \xi^a}=0\\
\label{FE-vphi}
\frac{\delta\LL}{\delta \varphi^a}&\equiv
\frac{\pd\LL}{\pd \varphi^a}-\d\, \frac{\pd\LL}{\pd \d \varphi^a}
-\pd_t\, \frac{\pd\LL}{\pd \dot \varphi^a}=0\\
\label{FE-phi}
\frac{\delta\LL}{\delta \phi^a}&\equiv
\frac{\pd\LL}{\pd \phi^a}+\d\, \frac{\pd\LL}{\pd \d \phi^a}
-\pd_t\, \frac{\pd\LL}{\pd \dot{\phi}^a}=0\, .
\end{align}
Alternatively, we may variate the total Lagrangian with respect to the 
gauge-invariant quantities, namely the incompatible
velocity $v^a$ and the coframe $\vt^a$:
\begin{align}
\frac{\delta\LL}{\delta v^a}&\equiv
\frac{\pd\LL}{\pd v^a}-\d\, \frac{\pd\LL}{\pd \d v^a}
-\pd_t\, \frac{\pd\LL}{\pd \dot v^a}=0\\
\frac{\delta\LL}{\delta \vt^a}&\equiv
\frac{\pd\LL}{\pd \vt^a}+\d\, \frac{\pd\LL}{\pd \d \vt^a}
-\pd_t\, \frac{\pd\LL}{\pd \dot{\vt}^a}=0\, ,
\end{align}
which can be expressed in the following form
\begin{align}
\label{FE1}
\frac{\delta\LL}{\delta v^a}&\equiv
\frac{\pd\LL}{\pd v^a}
-\d\, \frac{\pd\LL}{\pd I^a}
=0\\
\label{FE2}
\frac{\delta\LL}{\delta \vt^a}&\equiv
\frac{\pd\LL}{\pd \vt^a}+\d\, \frac{\pd\LL}{\pd T^a}
+\pd_t\, \frac{\pd\LL}{\pd I^a}=0\, .
\end{align}
These field equations of dislocation theory (\ref{FE1}) and (\ref{FE2})
may be expressed in terms of the response quantities (\ref{H-CR}),
(\ref{D-CR}), (\ref{E-CR}) and (\ref{p-CR}) in order to take the following from
\begin{align}
\label{YM-fe1}
\d D_a-\pi_a&=p^{\rm }_a\, \\
\label{YM-fe2}
\d H_a-\dot D_a-E_a&=\Sigma^{\rm }_a\, .
\end{align}
Eqs.~(\ref{YM-fe1}) and (\ref{YM-fe2}) are two Yang-Mills type field equations
with the momentum $(p_a+\pi_a)$ and the force stress $(\Sigma_a+E_a)$ as sources 
of the dislocation excitations $D_a$ and $H_a$, respectively.
The translational gauge fields themselves cause `gauge' sources $\pi_a$ and $E_a$,
thereby contributing to their own elastic sources $p_a$ and $\Sigma_a$.
Due to the complexity of the dislocation gauge field interaction, there
arise self-couplings which involve the dislocation momentum $\pi_a$ and the 
dislocation stress $E_a$. Thus, the field equations~(\ref{YM-fe1}) and
(\ref{YM-fe2}) are nonlinear. 
In addition, 
Eqs.~(\ref{YM-fe1}) and (\ref{YM-fe2}) constitute a closed system of 
12 independent field equations for the state quantities $v^a$ and $\vt^a$. 
Eq.~(\ref{YM-fe1}) can be interpreted as the 
equilibrium equation between the dislocation momentum flux and
the momenta (balance equation of momenta). 
In the static case, Eq.~(\ref{YM-fe1}) is vanishing.
Eq.~(\ref{YM-fe2}) is the equilibrium equation between the pseudomoment stress,
dislocation momentum flux and force stresses (balance equation of stresses).
From Eqs.~(\ref{YM-fe1}) and (\ref{YM-fe2}), 
we obtain the following conservation law:
\begin{align}
\label{eq2}
\dot p^{\rm }_a+\dot \pi_a+\d \big( \Sigma_a+E_a\big)=0\, .
\end{align}
Eq.~(\ref{eq2}) is nothing but the Euler-Lagrange equation~(\ref{FE-xi}) which
is the force equilibrium condition if 
dislocations are present (continuity equation of force stresses).
It determines the exchange of momentum and stress between the elastic and the
dislocation subsystems.
Observe that, in contrast to standard elasticity theory, the conserved quantities
are the total momentum and the total force stress of the system and not the elastic 
quantities themselves.
Linear solutions of the field equations~(\ref{YM-fe1}) and (\ref{YM-fe2}) for
moving dislocations are given by Lazar~\cite{Lazar09,Lazar10}.

In order to complete the framework of dislocation field theory, 
we introduce the {\it Peach-Koehler force} 3-form 
\begin{align}
\label{PKF}
f_a:=-\dot\pi_a-\d E_a
=\dot p_a+\d\Sigma_a
=\big(e_a\iner I^b\big)p^{ }_b
    +\big(e_a\iner T^b\big)\wedge\Sigma^{ }_b\, .
\end{align}
It represents the force density acting on dislocations. 
Eq.~(\ref{PKF}) is the dynamical form of the Peach-Kohler force 
which is analogous to the Lorentz force~\cite{HO01} in Maxwell's theory of 
electromagnetic fields.
The force density 3-form is a covector valued 3-form according to
\begin{align}
f_a=\frac{1}{3!}\, f_{aijk}\, \d x^i\wedge\d x^j\wedge \d x^k
\end{align}
and the dimension of the force density tensor is:
$[f_{aijk}]={\text{force}}/({\text{length}})^3$.

The {\it moment stress} 2-form $\tau_{ab}$ is related to the excitation $H_a$ 
according to
\begin{align}
\label{momstress}
\tau_{ab}:=\vt_{[a}\wedge H_{b]}\, 
\end{align}
with the dimension: $[\tau_{ab}]={\text{force}\times\text{length}}$.
In components, the moment stress 2-form reads
\begin{align}
\tau_{ab}=\frac{1}{2}\, \tau_{abij}\, \d x^i\wedge\d x^j
\end{align}
and the dimension of the moment stress tensor is:
$[\tau_{abij}]={\text{force}/\text{length}}$.
The formula~(\ref{momstress}) can be inverted
as follows:
\begin{align}
\label{last}
H_a=-2e_b \iner \tau_a{}^b + \frac{1}{2}\,  
\vartheta_a\wedge(e_b\iner e_c\iner\tau^{bc})\,\,.
\end{align}
In order to derive the {\it moment equilibrium}, it requires some algebra. We start with
the field equation (\ref{YM-fe2}) and compute the antisymmetric piece
of the total stress, use (\ref{momstress}), and find finally
\begin{equation}\label{momequi}
\d\tau_{a b}-T_{[ a}\wedge H_{ b]}+\vt_{[ a}\wedge\dot D_{ b]}
+\vt_{[ a}\wedge\left( E_{ b]}    +\Sigma_{b]}\right)=0\,.
\end{equation}
Apart from the terms $-T_{[a}\wedge H_{b]}$ and 
$\vt_{[ a}\wedge\dot D_{ b]}$, this is exactly the expected law known from
continuum mechanics.

\subsection{Quadratic gauge field Lagrangian} 
Now we may specify {\it constitutive laws.} 
In order to give concrete expressions for the excitations, 
we have to specify the constitutive relations between field strengths
$(T^a,I^a)$ and excitations $(H_a,D_a)$. 
For a local, linear, isotropic continuum we have 
\begin{align}
\label{const_iso}
H_a=g_{ab}\,\hodge\!\sum_{I=1}^{3}a_{I}\,^{(I)}T^b\, ,
\end{align}
wherein $^{(I)}T_a$ are the irreducible pieces (\ref{tentor}),
(\ref{trator}), and (\ref{axitor}) of the torsion and $a_1$, $a_2$,
and $a_3$ are constitutive moduli 
which have the dimension: $[a_I]=\text{force}$.
$\hodge T^a$ is dimensionless.
We may decompose the
torsion into three $SO(3)$-irreducible pieces according to
$T^a=\,^{(1)}T^a+\,^{(2)}T^a+\,^{(3)}T^a$ with the number of
independent components $9=5\oplus 3\oplus 1$. These three pieces
(tensor piece, trace- and axial-vector pieces of the torsion) are defined by
\begin{alignat}{2}
\label{tentor}
^{(1)}T^a&:=T^a-\,^{(2)}T^a-\,^{(3)}T^a
&&\qquad \text{(tentor)},\\
\label{trator}
^{(2)}T^a&:=\frac{1}{2}\,\vartheta^a\wedge\big(e_b\iner T^b\big)
&&\qquad\text{(trator)},\\
\label{axitor}
^{(3)}T_a&:=\frac{1}{3}\,e_a\iner\big(\vartheta^b\wedge T_b\big)
&&\qquad\text{(axitor)}\, .
\end{alignat}
In addition, we have for the local, linear, isotropic continuum the constitutive
relation between $D_a$ and $I^a$:
\begin{align}
\label{const_iso-D}
D_a=g_{ab}\,\hodge\!\sum_{I=1}^{3}f_{I}\,^{(I)}I^b\, ,
\end{align}
wherein $^{(I)}I_a$ are the irreducible pieces (\ref{ten}),
(\ref{skew}), and (\ref{tra}) of the dislocation  current and
$[\hodge I^a]=(\text{length})^2/\text{time}$. $f_1$, $f_2$,
and $f_3$ are constitutive moduli which have the dimension: 
$[f_I]=\text{mass}/\text{length}
\stackrel{\rm SI}{=}\text{kg}/\text{m}$.
The three $SO(3)$-irreducible pieces 
$I^a=\,^{(1)}I^a+\,^{(2)}I^a+\,^{(3)}I^a$ with the number of
independent components $9=5\oplus 3\oplus 1$ are given by
\begin{alignat}{2}
\label{ten}
^{(1)}I^a&:=I^a-\,^{(2)}I^a-\,^{(3)}I^a
&&\qquad \text{(symmetric and traceless)},\\
\label{skew}
^{(2)}I_a&:=\frac{1}{2}\,e_a\iner\big(\vartheta^b\wedge I_b\big)
&&\qquad\text{(antisymmetric)},\\
\label{tra}
^{(3)}I^a&:=\frac{1}{3}\,\vartheta^a \big(e_b\iner I^b\big)
&&\qquad\text{(trace)}\, .
\end{alignat}

Thus, for a linear continuum,
the dislocation Lagrangian has the bilinear form
\begin{align}
\label{L_disl1}
{\cal L}_{\rm disl}=\frac{1}{2}\,I^a\wedge D_a-\frac{1}{2}\,T^a\wedge H_a\, .
\end{align}
The pure {\it dislocation energy} is defined as the Hamiltonian of the
dislocation system
\begin{align}
\label{E_disl}
{\cal E}_{\rm disl}:=I^a\wedge D_a-{\cal L}_{\rm disl}
                   =\frac{1}{2}\,I^a\wedge D_a+\frac{1}{2}\,T^a\wedge H_a\, .
\end{align}
More physically, we can interpret ${\cal E}_{\rm disl}$ as the dislocation 
core energy. Moreover, we see in Eqs.~(\ref{L_disl1}) and (\ref{E_disl})
that 
the first term plays the role of the kinetic dislocation energy and 
the second one can be identified 
with the potential energy of dislocations.
Thus, the excitation $D_a$ is a kind of `momentum' and $I^a$ plays 
the role of a generalized `velocity' of the dislocation motion.

\section{Discussion and Conclusion}
\begin{table}[t]
\begin{tabular}{ll}
Maxwell field theory & Dislocation field theory\\
\hline
\hline
$B$ - magnetic field strength & $T^a$ - dislocation density  \\
$E$ - electric field strength & $I^a$ - dislocation current\\
$H$ - magnetic excitation     & $H_a$ - pseudomoment stress \\
$D$ - electric excitation     & $D_a$ - dislocation momentum flux\\
$A$ - magnetic potential 1-form        & $\phi^a$ - dislocation potential 1-form\\
$\varphi$ -  potential 0-form & $\varphi^a$ - dislocation potential 0-form\\
$f$ - gauge function      & $\xi^a$ - deformation mapping\\ 
$B=\d A$,\quad $A^\prime=A+\d f$ & $T^a=\d\phi^a$,\quad $\phi^{a\prime}\equiv\vartheta^a=\phi^a+\d\xi^a$\\
$E=\d\varphi-\dot A$,\quad $\varphi^\prime=\varphi+\dot f$ & $I^a=\d\varphi^a-\dot\phi^a$,\quad
$\varphi^{a\prime}\equiv v^a=\varphi^a+\dot\xi^a$\\                               
$\Phi$ - magnetic flux of magnetic vortices & $b^a$ - Burgers vector of dislocations\\
$\int_{S} B=\Phi$ & $\int_{S} T^a=b^a$\\
$\rho$ - electric charge density & $p^{\rm T}_a=p_a+\pi_a$ - total momentum density \\
$j$ - electric current     & $\Sigma^{\rm T}_a=\Sigma_a+E_a$ - total force stress \\
\hline
magnetic field closed: & continuity equation of dislocation density:\\
$\d B=0$        & $\d T^a=0$\\
Faraday law: & continuity equation of dislocation current:\\
$\d E+\dot B=0$ & $\d I^a+\dot T^a=0$\\
Gauss law: & continuity equation of dislocation moment flux: \\
$\d D=\rho$     & $\d D_a=p^{\rm T}_a$\\
Oersted-Amp{\`e}re law: & continuity equation of moment stress:  \\
$\d H-\dot D=j$ & $\d H_a-\dot D_a=\Sigma^{\rm T}_a$\\
continuity equation of current:& continuity equation of force stress:\\
$\dot\rho+\d j=0$ & $\dot p^{\rm T}_a+\d\Sigma^{\rm T}_a=0$\\
constitutive laws:& constitutive laws:\\
$H=H(B)$, $D=D(E)$ & $H_a=H_a(T^b)$, $D_a=D_a(I^b)$\\
Lorentz force density: & Peach-Koehler force density: \\
$f_a=(e_a\iner E)\rho+(e_a\iner B)\wedge j$
&$f_a=(e_a\iner I^b) p_b+(e_a\iner T^b)\wedge \Sigma_b$ \\
\hline
\end{tabular}
\caption{The correspondence between Maxwell's theory and dislocation field
  theory.}
\label{tab1}
\end{table}

Based on the $T(3)$-gauge theory, 
we have proposed a dynamical field theory of dislocations. 
We have used the concepts of field strengths, excitations and constitutive laws
analogical to the electromagnetic field theory
in order to obtain a closed field theory.
All dislocation field quantities can be described by ${\mathbb R}^3$-valued
exterior differential forms. The translation field strengths are 
even (or polar) differential forms and the excitations 
(stresses and momenta) are odd (or axial) forms.
We have shown that the excitations to the
dislocation density and dislocation current are necessary for a 
realistic physical dislocation field theory.
Moreover, we have demonstrated how the excitations have to fit into the 
Maxwell-type field equations in contrast to~\cite{goleb} who claimed 
that, there are no analogues to the second pair of the Maxwell equations in 
dislocation theory. 
A review of the corresponding electromagnetic and dislocation quantities 
is given in Table~\ref{tab1}.
The gauge theory of dislocations is a closed field theory.
However, there are important distinctions.
In the Maxwell theory the field quantities are scalar-valued forms and in the 
dislocation field theory all field quantities are vector-valued (or
covector-valued) forms. Due to self-couplings the field equations for dislocations  
are nonlinear. The electromagnetic current $j$ depends on an exterior material
field,
while the stress $(E_a+\Sigma_a)$ is interior and it depends on the coframe
(distortion) itself. 

It is known that dislocations in crystals move in two different modes, called
glide (conservative motion) and climb (non-conservative motion). 
For instance, the volume and the mass density, respectively, of the 
crystal is not changed by the gliding of dislocations.
On the other hand, climbing dislocations interchange with point defects 
such as vacancies and/or interstitials. 
Additionally, if dislocations cut each other, they build networks of dislocations.
In our dynamical dislocation theory, we have neglected dissipation (friction and radiation damping) and the 
interaction with point defects. 
In order to take into account the energy dissipated and 
converted into heat one can use a Lagrangian extended by a dissipation 
function (see, e.g., \cite{LL}). 

To sum up, the state quantities in dislocation dynamics are the physical velocity $v^a$, the
elastic distortion $\vt^a$, the dislocation density $T^a$ and 
the dislocation current $I^a$. 
In a dislocation field theory based on these state quantities one finds 
the Euler-Lagrange equations and the response quantities as we have given 
in this paper. In addition, one can combine such a dislocation theory with the
so-called `multiplicative decomposition' \cite{Bilby57,Kroener60,Lee}
which is widely used and accepted in 
engineering science. Nevertheless, the geometrical or field theoretical arena 
of a dislocation field theory is the gauge theory of the three-dimensional 
translation group. It is just a consequence from the fact that dislocations break 
locally the translation symmetry in the crystal and that the dislocation density 
tensor is nothing but a realization of Cartan's torsion tensor~\cite{Cartan1922,Cartan} in 
three dimensions what was originally found by Kondo~\cite{Kondo} 
(see also \cite{Kroener60}).

\section*{Acknowledgement}
The author was supported by an Emmy-Noether grant of the 
Deutsche Forschungsgemeinschaft (Grant No. La1974/1-2,3). 

\end{document}